\documentclass[a4paper,12pt]{article}
\usepackage{epsfig}
\usepackage{amssymb}
\begin{document}
\thispagestyle{empty}

%____________________________________________
%\newcommand{\JP}    {jet probability}
\newcommand{\etal}  {{\it{et al.}}}  % must have all the {}
\def\Journal#1#2#3#4{{#1} {\bf #2}, #3 (#4)}
\def\PRD{Phys.\ Rev.\ D}
\def\NIMA{Nucl.\ Instrum.\ Methods A}
\def\PRL{Phys.\ Rev.\ Lett.\ }
\def\PLB{Phys.\ Lett.\ B}
\def\EPJ{Eur.\ Phys.\ J}
\def\IEEETNS{IEEE Trans.\ Nucl.\ Sci.\ }
\def\CPCD{Comput.\ Phys.\ Commun.\ }
%____________________________________________

%\hfill {\LARGE\bf DRAFT}
%\smallskip

%\hfill {\large\bf \today}
\bigskip
%\bigskip

%{\LARGE\bf
{\Large\bf
\begin{center}
%Inhomogeneities in the model with gravitational stringlike compact objects
The formation of inhomogeneities in gravitational string like model
\end{center}
}
\vspace{0.1 cm}
%\vspace*{\fill}

\begin{center}
{ G.A. Kozlov  }
\end{center}
%\vspace{1 cm}
%\vspace*{\fill}
%\date
\begin{center}
\noindent
 { Bogolyubov Laboratory of Theoretical Physics\\
 Joint Institute for Nuclear Research,\\
 Joliot Curie st., 6, Dubna, Moscow region, 141980 Russia  }
\end{center}
%\vspace*{\fill}
\vspace{0.1 cm}

 %\section*{Abstract}
 \begin{abstract}
 \noindent
 {The investigation of the inhomogeneities in modern inflationary Universe scenarios is related, in particular, with the study of the role played by scalar fields in cosmological evolution. 
 We present the model described by one of the extensions of the Abelian-Higgs theory, where the scale-invariant approach to the  Standard Model (SM) is achieved by introducing the complex scalar fields and the vector fields in the hidden (dark) sector, where  the latter and the SM are minimally coupled to gravity.
 The formation of the cosmological compact objects intuitively is explained in terms of the topological line defects - the string like tubes.
 The strings can get an energetically preferable configuration corresponding to the magnetic flux quantum inside an area defined by the mass of dark vector field.}

%\vspace*{\fill}

\end {abstract}

%\vspace*{\fill}

%\newpage
%\tableofcontents

%\newpage
%\section{Introduction}

\bigskip

%\section{Introduction}

The investigation of an emergence and the structure of inhomogeneities in modern inflationary Universe scenarios is based, in particular, with the study of the role played by scalar fields in cosmological evolution [1]. 
Within the quantitative research, 
%e.g., in the form of string-like vortex flows in the early Universe, may also shed light to additional suggestions to explain the baryon asymmetry [7]. 
there could have appeared a hope to get the field parameter space where the emergence of inhomogeneities  may be really generated in the natural observations which are based on the models beyond the Standard Model (SM).  
In the case of a complex scalar field, the growth of the gravity instability may be caused by deviations from the spatial uniformity related to the presence of strings  [2-9]. 
One can suppose the appearance of the cosmological compact objects (CCO)  considered as a  stationary configuration of new particle species 
%the scalar and vector fields 
in terms of the flux tube (FT) under the impact of the gravity (GR). 
These new degrees of freedom, the bosons, suggested as a component of dark matter (DM), can form self-gravitating structures which may describe well the core of DM halos when the bosons are ultralight with masses $\sim O(10^{-22}$ eV).
The stability of compact objects  can arise naturally due to GR collapse of the constituents in the final end-state of the CCO. 
The absence of direct and indirect detection of the compact objects leaves the properties of the fields inside and the mechanism of the CCO formation still largely unknown.
We present the model based on a minimal extension of the SM with GR that contains no dimensional parameters in the action and is, therefore, scale-invariant at the classical level. In the model, all scales are induced by the spontaneous breakdown of scale invariance (SI). To that end, we suppose that in the formation of the CCO the main contribution comes from the strong DM sector that is minimally coupled to GR in the sense of dynamical dimensional constants
%, i.e., the Planck mass $M_{Pl}\simeq 1.2\cdot 10^{19}$ GeV  may be the dynamical quantity 
[10].
We use the one of the extensions of the Abelian-Higgs model [11-14] where  
%the $U(1)$ symmetry is supposed to be spontaneously broken, 
the string like excitations may be the key components of the field configuration of the compact FT under the impact  of the scalar DM fields squeezing the flux into a tube with the radius $r$. 

Let us start with the classical scheme containing the scalar and vector sectors with GR interactions inside the small cosmological "structure", the galaxy halo. 
In the model, the SI is spontaneously broken at the energy scale $\sim\phi_0$. We refer to this new particle species as the DM scalar.
% and  $\phi_0$ is its decay constant. 
The symmetry to be explicitly broken due to DM scalars being coupled to a new gauge force that becomes strong at low energies.
The Lagrangian density  (LD) is
\begin{equation}
\label{e141}
 \frac{L}{\sqrt{-g}} = \frac{1}{2} M_{Pl}^2\mathcal{R} + L_{[\lambda\rightarrow 0]}^{SM} - V(H, \phi) -\Lambda_{cosm}, 
%\frac{L}{\sqrt{-g}} = \frac{\mathcal{R}}{16\pi}M_{Pl}^2  - \frac{1}{2} g^{\mu\nu}\partial_\mu\phi\,\partial_\nu\phi^\star + L_D,  
\end{equation}
where the first term is the Einstein-Hilbert action for the gravity with the Planck mass $M_{Pl}\simeq 1.2\cdot 10^{19}$ GeV; $\mathcal{R}$ is the Ricci scalar for background metric $g_{\mu\nu}$. 
The second term is the SM-like LD containing no self-interacting fields (the self-coupling constant $\lambda\rightarrow 0$).  
The DM scalar potential arises below some strong coupling scale $\Lambda_{sp} = M_{Pl}/\sqrt {N_{sp}}$, $N_{sp}$ being the number of the new scalar species.
The scale $\Lambda_{sp}$ would be arbitrary small, $\Lambda_{sp} << \phi_0 \leq M_{Pl}$ whenever we suppose a parametrically large $N_{sp}$. 
Around the scale $\Lambda_{sp}$, the particle species turn into a tower of hidden DM scalars. 
This should be happened whenever we get an infinite tower of states  $\phi (x) = \sum_{k=1}^{N_{sp}} c_k \varphi_k (x)$ for scalar fields $\varphi_k (x)$ becoming light of different masses $\mu_k$. The coefficient $c_k$ plays the role of conformal regulator depending on the size of conformal mass parameter $\Delta^2$ relevant to each value $\mu_k^2 = \mu_0^2 + (\Delta^2/N_{sp})k$ $(k =1,2, ..., N_{sp})$, while $c_k\varphi_k (x)$ is the conformal element in the tower operator $\phi (x)$ [15].
The potential $V(H, \phi)$ in (\ref{e141}) has the usual symmetry breaking form
\begin{equation}
\label{e1411}
V(H,\phi) = \lambda\left ( H^+H - \frac{\alpha}{2\lambda} {\vert \phi\vert}^2\right )^2
\end{equation}
for the SM Higgs doublet $H^+H$ and the complex scalar fields $\phi$. In (\ref{e141}),  $\Lambda_{cosm}$ is the cosmological constant (CC). Tree-level stability and positivity of (\ref{e1411}) both require $\lambda > 0$ and  the constant $\alpha > 0$. The SM sector and the DM scalar one decouple when $\alpha \rightarrow 0$. Communication of the $\phi$-sector with the SM proceeds via the quartic term $\sim H^+H{\vert \phi\vert}^2$, by which the Higgs field becomes a portal to new scalar sector.
%with the coupling constant $\alpha$. 
The classical SI is violated by $M_{Pl}$ and $\Lambda_{cosm}$ in (\ref{e141}). 
We use the SI scheme, where the dimensional constants in (\ref{e141}) be dynamical, i.e. replace them by the fields [10]. The LD becomes 
\begin{equation}
\label{e1412}
%\frac{L}{\sqrt{-g}} = \left (\frac{1}{2}\zeta_\phi {\vert \phi\vert}^2 + \zeta_h H^+H +\frac{1}{2}\zeta_B B_\mu B^\mu\right )\mathcal{R} - \frac{1}{4} B_{\mu\nu}B^{\mu\nu} + {\vert \left %(\partial_\mu + i g_D B_\mu\right )\phi\vert}^2 - V_f (H, \phi).
\frac{L_{SI}}{\sqrt{-g}} = \left (\frac{\zeta_\phi}{2} {\vert \phi\vert}^2 + \zeta_h {\vert H\vert}^2 +\frac{\zeta_B}{2} B_\mu^2 \right )\mathcal{R} - \frac{1}{4} B_{\mu\nu}^2 + {\vert \left (\partial_\mu + i g B_\mu\right )\phi\vert}^2 - V_f (H, \phi).
\end{equation}
Here, $\zeta_\phi$, $\zeta_h$  and $\zeta_B$ are real positive parameters. 
%The dark $\phi$ field that provides the $B_\mu$ field its mass is a complex scalar charged $B_\mu$ with gauge coupling $d_A$ and 
The model may realise the cold DM $B_\mu$ with $B_{\mu\nu} = \partial_\mu B_\nu - \partial_\nu B_\mu$ and $g$ being the gauge coupling, which controls the strength of backreaction of dark field $B_\mu$ onto dark field $\phi$, is a free parameter and can be tuned small enough to avoid the defect formation. 
 In the broken phase, the field $B_\mu$ acquires a mass $m_B \sim g\phi_0$ and contributes $\sim g^2 {\vert\phi\vert}^2 B_\mu^2$ to the effective potential with $\phi_0$ being the classical ground state of $\phi$. This contribution coincides with the local energy density of the DM $\rho_{DM}$ in the non-relativistic limit. At the Hubble rate $H_0 = 10^{-33}$ eV, the point $B_\mu B^\mu$ is replaced by $\sim\rho_{DM}/m^2_B$. 
%For the FT formation understanding, the cold DM vector sector can be approximated as a constant  homogeneous magnetic field (see, e.g., [15] and the refs. therein). 
%For the potential energy with the complex scalar fields at the stage where the mass of the scalar field exceeds $H$, the formation of the structure of inhomogeneities are possible [8]. 
%The (\ref{e1412}) is of the type of the LD that has the flux-tube like solution in the sense of the vortex [10-12].
We introduce the small quantity $ f << 1$ as the degree of the flatness  in the potential $V_f (H,\phi) = V(H,\phi) + f {\vert \phi\vert}^4$, where the last term is, in fact, associated with the CC, dark energy (DE).
The mixing of the Higgs with the DM scalar field, setting the scale for the breaking of SI, can generate a second, cosmologically long-lived vacuum with small positive $f$.
For the potential energy with the complex scalar fields at the stage where the mass of the scalar field exceeds $H_0$, the formation of the structure of inhomogeneities is possible [1]. 
The (\ref{e1412}) is of the type of the LD that has the FT-like  solution. 

The departure from SI is related, in particular,  with the one loop  contributions to the potential with complex self-interaction scalar field. We also point out that the vacuum of the theory, in general, depends on the temperature of the Universe.
Any interactions between the fields should lead to thermal equilibrium, and in case of large occupation number of (super)light particles - to formation of a Bose-Einstein condensate.
% which break classical SI. 
There are new updated terms that appear in (\ref{e141}) and (\ref{e1411}) in this case,
%\begin{equation}
%\label{e1413}
 $\Lambda_{cosm}\rightarrow\Lambda_{cosm}^\prime = \Lambda_{cosm} (1+ \delta_r), $
 %\,\,\, \sim f{\vert\phi\vert}^4 +\frac{\alpha^2}{8\lambda} {\vert\phi\vert}^2 T^2,
 %\end{equation}
 and $\sim \left [f{\vert\phi\vert}^2 +\alpha^2 T^2/{(8\lambda)}\right ] {\vert\phi\vert}^2$,
where $\delta_r\sim\rho_r/\Lambda_{cosm} <1$.
% $\rho_T = (\pi^2/90) T^4$ denotes the radiation-dominated energy-density of the Universe at the temperature $T$ (the thermal bath). 
In the early Universe, the radiation-dominated energy-density at the temperature $T$ (the thermal bath), $\rho_r\sim T^4g_\star$, is the constant pressure that may capture the interactions between the SM and the DM particles in the plasma that generates an effective potential 
%$ V(H,\phi;T)$ 
with $T$-dependent terms [16-18]. For simplicity, we take the number of relativistic degrees of freedom $g_\star$ as a constant. The effect of SI breaking is negligible if $T <<  (\sqrt{\lambda f}/\alpha)T_{Pl}$ for  ${\vert\phi\vert}^2$ up to maximal value 
$\sim M_{Pl}^2$ (corresponding to DE models proposed in terms of Quintessence 
described by the scalar fields as the dynamical DE minimally coupled to GR [19]) with  the Planck temperature $T_{Pl}\sim 10^{32}$K; $f << \alpha << 1$ and $\lambda\leq O(1)$ as it corresponds to the self-interaction of the Higgs field.
%$T^2 << (8\lambda/\alpha^2) f {\vert\phi\vert}^2$. 
The contribution of the Higgs and the new scalar field (setting the scale for breaking the SI) with the impact by GR, results in the ground state $v^2 = \lambda^{-1} (\zeta_h \mathcal{R} + \alpha\phi_0^2)$ with corresponding family of classical ground states given by $H^+H = v^2/2$ and ${\vert\phi\vert}^2 = \phi_0^2$. The scalar constant curvature is $\mathcal{R} = 4 f \phi_0^2\left (1 - \delta_{DM}^T\right)/A_{\phi h}$, where $\delta_{DM}^T = \rho_{DM} \left (1 - T^2/T^2_\star\right)/(4 f \phi_0^4)$, $A_{\phi h} = \zeta_\phi +(\alpha/\lambda)\zeta_h$. 
The SI can be spontaneously broken in the absence  of the gravity  when the temperature achieves the value $T_\star^{SI} = T_\star\sqrt {1 - 1/\delta_{DM}^{T=0}}$ with the characteristic temperature  $T_\star = 2\sqrt {\lambda\rho_{DM}}/(\alpha \phi_0)$. 
%In this case, $f < \rho_{DM}/(4 \phi_0^4)\sim \zeta_\phi^2\times 10^{-118}$. 
The theory with not containing 
$\mathcal{R}$ can not explain the observed accelerated expansion of the Universe without introduction the DE component. 
Thus, we consider the theory in the range $T\in [0, T_\star^{SI})$. 

%In the model, the solutions with $\phi_0\neq 0$ spontaneously break SI. 
At the classical level, all scales are induced and proportional to $\phi_0$, in particular, 
$$M_{Pl} = {\left [A_{\phi h} +  \frac{4f\, \zeta_h^2}{\lambda A_{\phi h }} \left (1 - \delta_{DM}^T\right ) + \frac {2\zeta_B\delta_{DM}^{T=0}}{g^2} f \right ]}^{1/2} \phi_0. $$
The $\phi_0$ lies within the upper limit in the scale range $H_{inf} < \Lambda_{sp} < T_{max} < \phi_0 < M_{Pl}$ if $A_{\phi h} \geq 1$ with $H_{inf}$ being the inflationary Hubble rate, $T_{max}$ is the maximal radiation temperature, $T_{max}\sim g_\star^{-1/4}\sqrt {H_{inf} M_{Pl}} >> T_\star$.
The hierarchy rates between $\Lambda_{cosm}^\prime$, $M_{Pl}$ and the electroweak (EW) scale $M_{EW}$ are:
\begin{equation}
\label{e1461}
 \frac{\Lambda_{cosm}^\prime}{M_{Pl}^4}\simeq \frac{4f}{A_{\phi h}^2}\left (1 - \delta_{DM}^T\right )\left ( 1- \Delta_B\right) + \frac{\pi^2}{90} T^4\times 10^{-76} \frac{1}{GeV^{4}}  \sim O\left ( 10^{-120}\right ), 
 \end{equation}
%$$   \frac{M_{EW}^2}{M_{Pl}^2}\sim \frac{\alpha}{\lambda \zeta_\phi}\left ( 1- \Delta_B\right)\sim O\left ( 10^{-34}\right ), $$
\begin{equation}
\label{e1462}
 \frac{M_{EW}^2}{M_{Pl}^2}\simeq \frac{1- \Delta_B}{\lambda A_{\phi h}}\left [\frac{4 f \zeta_h}{A_{\phi h}}\left (1 - \delta_{DM}^T\right ) + \alpha\right ]\sim O\left ( 10^{-34}\right ), 
 \end{equation}
where $f < \zeta_\phi^2\times 10^{-118}$, $\zeta_\phi\leq A_{\phi h}$, $\alpha \leq \lambda\zeta_\phi\times 10^{-34}$; $1 - \Delta_B \simeq \zeta_\phi  \phi^2_0/M^2_{Pl}$ and $\Delta_B = \zeta_B\rho_{DM}/\left (m^2_B M^2_{Pl}\right )$.
% and $f > \zeta_\phi^2\times 10^{-118}$. 
 For model-independent lower bounds on DM mass in the local Universe $ > 10^{-19}$ eV [20] and $ > 2.2\times 10^{-21}$ eV [21], the $\Delta_B$ is restricted by $\Delta_B < \zeta_B\times \left (10^{-24} - 10^{-20}\right )$ for the local energy density $\rho_{DM} = \rho_\odot = 0.43$ GeV/${cm^3}$ of the Solar system [22].
% where the DM thermal contribution gives $f\delta_{DM}^T\simeq \zeta_\phi^2 \left (1 - T^2/T^2_c\right)\times 10^{-118}$. 
Using (\ref{e1462}), the characteristic temperature is estimated at the level of $T_\star\simeq (1/\sqrt\zeta_\phi)\times 10^7$K.
%If we use the largest energy-density $\rho_{DM} = 5\times 10^9$ GeV/${cm^3}$ that one would expect for stars near the Galactic centre 
In the scenarios with the DM spike  for stars near the Galactic centre [23] with $\rho_{DM} = 5\times 10^9$ GeV/${cm^3}$,
%the upper limit for $\Delta_B$ would be as of ten orders higher than that if use $\rho_\odot$, however 
the characteristic temperature can reach $T_\star\sim 10^{12}$K which is extremely high value assuming the very early Universe at the border of the quark-gluon formation stage. 
For $T\sim T_0^\gamma \sim 2.7$K (the Cosmic Microwave Background radiation $T$), the second term in (\ref{e1461}) contributes with about $\sim 10^{-128}$, thus leading to the fact that the main impact to the hierarchy rate (\ref{e1461}) comes from the first term with the appropriate parameters.
If the scalar species are associated with the candidates to DM and being cosmologically long-lived particles, the number of the species $N_{sp}$ can be estimated. In this case, the lightest state $\phi$ could decay into two gravitons ($GG$) with the effective couplings between the $\phi$ and the quadratic curvature, the square of Riemann tensor, $\sim c_{\phi GG}(\phi/\Lambda_{sp})R_{\mu\nu\rho\sigma}R^{\mu\nu\rho\sigma}$ [24]. Looking through the decay width 
$$\Gamma (\phi\rightarrow GG)\simeq \frac{c_{\phi GG}^2}{2\pi} \left (\frac{m}{M_{Pl}}\right )^7 \frac{N_{sp}}{\tau_U}\times 10^{-23} GeV^2,$$
the lifetime of $\phi$ against the lifetime of the Universe $\tau_U\sim 10^{17}$ sec can be long enough if the mass $m$ of the $\phi$ boson $m < 10^{22} GeV/(c_{\phi GG}^2 N_{sp})$ from which one finds $N_{sp} < 10^{52}$ if the free parameter $c_{\phi GG}\sim O(1)$ and $m > 10^{-21}$ eV [21]. 

The DM field $B_\mu$ is 
\begin{equation}
\label{e147}
%g  B_\mu (x)  = - \partial_\mu\eta (x) + \frac{1}{2 g}\frac{\alpha}{\lambda} \frac{k_\mu}{{\vert H\vert}^2} \left [1 + O\left (\zeta_\phi\zeta_h\frac{\phi_0^2}{{\vert H\vert}^2}\times %10^{-120}\right )\right ],
g  B_\mu (x)  = \frac{1}{1 + \delta_\varphi}\left [- \partial_\mu\eta (x) +  \frac{k_\mu (x)}{2 \,g\, \varphi^2 (x)}\right ],
%{{\vert H\vert}^2} \left [1 + O\left (\zeta_\phi\zeta_h\frac{\phi_0^2}{{\vert H\vert}^2}\times 10^{-120}\right )\right ],
\end{equation}
obeying the Eq.
\begin{equation}
\label{e146}
k_\mu \equiv \partial^\nu B_{\mu\nu} = \left (2 g^2 {\vert\phi\vert}^2 + \zeta_B \mathcal{R}\right )B_\mu - i g \left (\phi^\star\partial_\mu \phi - \phi\,\partial_{\mu}\phi^\star\right ),
\end{equation}
%(\ref{e146}), 
where the polar decomposition of the scalar field is used, $\phi (x) = \varphi (x) e^{i\eta (x)}$, and $\delta_\varphi = \zeta_B \mathcal{R}/(2 g^2 \varphi^2)$.
%The Higgs boson contribution enters through $\varphi ^2 = \alpha^{-1} \left (2\lambda {\vert  H\vert}^2 - \zeta_h \mathcal{R}\right )$.
The magnetic flux associated with $B_\mu$ can be quantised if the proper boundary conditions of Eq. (\ref{e146}) are known. The flux is 
\begin{equation}
\label{e149}
\Phi = \int B_{\mu\nu} \,d\sigma^{\mu\nu} = \oint  B_{\mu}(x)\,dx^\mu,
\end{equation}
where $\sigma^{\mu\nu}$ is a two-dimensional surface element $\sim \Delta x^\mu\Delta x^\nu$ $(\mu\neq\nu)$ in Minkowski space.
% in $S(\Re^4)$. 
The flux (\ref{e149}) guarantees the non-observability of the FT when the  scale and gauge symmetries are not broken.
At the cosmological scale, the contribution  to the flux (\ref{e149}) from the second term in the R.H.S. of (\ref{e147}) is negligible 
%because $\alpha << \lambda \leq O(1)$, and 
if we integrate out over a large closed loop, where  the generating current $k_\mu$ is vanished.  
Here, the assumption that the phase $\eta (x)$ in (\ref{e147}) is that the field $\phi(x)$ should be a single valued, is the result of variation with the winding integer number $n$ (the "topological charge" of the flux) by $2\pi n$ over a large closed loop. Hence, the flux (\ref{e149}) is quantised as $ -2\pi n/g$, resulting from the formal feature, that $\Phi$ should not be defined more precisely than that of $2\pi$ times $n$. The  flux is squeezed into compact object, a stringlike tube (the topological line defects - Nielsen-Olesen (NO) strings [11]), caused by the condensate of the scalar field, where each string has a quantised magnetic flux unit $\Phi \sim -2\pi/g$. The strings interact with the background field $B_\mu$, converting DM into string energy.

The cosmological compact objects are identified in terms of FTs within
%which are distributed inside the object under the requirement of translational invariance within the scalar condensate configuration. The latter is governed by 
the field equations 
% $2 f \left (\phi_0^2 - {\vert\phi\vert}^2 \right ) \phi = {\left (\partial_\mu + i g_D B_\mu\right )}^2 \phi$ 
%(\ref{e145}) 
%$$2 f \left [\phi_0^2 \left (1 - \delta_{DM}^T\right ) - \frac{\alpha^2}{16\lambda f} T^2 - {\vert\phi\vert}^2 \right ] \phi = {\left (\partial_\mu + i g B_\mu\right )}^2 \phi$$
$$\left [\frac{1}{2}\zeta_\phi\mathcal{R} +\alpha \left (H^+H\right) -\left (\frac{\alpha^2}{2\lambda} +2f\right ){\vert\phi\vert}^2 - \frac{\alpha^2}{8\lambda f} T^2 \right ] \phi = {\left (\partial_\mu + i g B_\mu\right )}^2 \phi$$
and (\ref{e146}) with an appropriate coordinate symmetry. For the FT as the field topological object, the cylindrical symmetry with the coordinate $r$  (the radial distance from the center of the FT) is most suitable, where $\varphi = \varphi (r)$ and $B_\mu\rightarrow \vec B = [\bar B (r) / r]\vec e_{\theta_{az}}$ with $\vec e_{\theta_{az}}$ being the unit vector. The field phase $\eta = n \theta_{az}$, where $\theta_{az}$ is the azimuth angle around $z$ axis, while $\vec\nabla \eta = (n/r)\vec e_{\theta_{az}}$. The rotation of $B_\mu$ field gives the electric field $\vec E = E_z(r) \vec e_z$ with the unit vector $\vec e_z$ along the $z$ axis.
%In the early stage of the BS formation, $\varphi (r)\sim 0$, $\bar B (r)\sim 0$ as the FT radius $r\rightarrow 0$. 
The essential point  is  the singularity at the center of the FT which comes from the quantisation  condition $\vec \nabla\times \vec\nabla \eta= (2\pi n) \delta (x)\delta (y) \vec e_z$, where the $\delta$-functions just point to the center of the FT. 
Since the phase $\eta$ is an arbitrary function with $n$, the latter is free to grow up as winding number. When it is happened, the flux has to grow as well as the characteristic size of the FT.
The profiles of  $\bar B(r)$ and $E_z (r)$  with the radial distance $r\sim\xi$, the penetration depth, and the coherence length $l\sim m_\varphi^{-1}$ of the $\varphi$-field, depend on the ratio, $\sim \alpha/(g \sqrt \lambda)$,  between the $\varphi$-scalar mass $m_\varphi = \sqrt{2/\lambda}\, \alpha\,\phi_0\simeq  \sqrt{2/(\zeta_\phi \lambda)}\, \alpha\, M_{Pl} \sim\sqrt\zeta_\phi\times 10^{-6}$ eV 
%($\zeta_\phi\sim O(10^{-3}), [8])$ 
and the mass of the vector field $m_B  = \sqrt 2\, g \,\phi_0\simeq \sqrt {2/\zeta_\phi}\, g\, M_{Pl}$. 
%where
%Here, $\alpha\sim \lambda\zeta_\phi (1- \Delta_B)^{-1}\times 10^{-34}$ with $\Delta_B\sim \zeta_B m_B^{-2}\times 10^{- 62} eV^2 < \zeta_B\times (10^{-24} - 10^{-20})$, and 
%the dark gauge coupling constant should be as $g > \zeta_\phi\times 10^{-34}$.)
% From now on we define $\phi_0^2\simeq \zeta_\phi^{-1} (1 - \Delta_B)M_{Pl}^2$.
There can be the minimal value $\zeta_\phi^{min}$ below which no strings (as well as the flux $\Phi$) could be produced ($r(\zeta_\phi = 0) = 0)$.

In the Abelian-Higgs theory, the vacuum is characterised by a ring of degenerate vacua allowing for NO strings, where each string has a quantised magnetic flux unit $\sim -2\pi/g$ with the preference to be confined to a topological defect. For the FT formation understanding, the cold DM  is approximated as an almost constant  homogeneous magnetic field [25].
%$g\mathcal{B}\sim g\rho_{DM}^{1/2}$ [24].
 The magnetic field background at zero temperature can modify the physical properties of the vacuum from which, in general, the flux tubes would interact with each other.  The field interactions of the model can provide us with the source of vacuum instability at the critical magnetic field.
In the broken phase, where the magnetic field $\mathcal{B}$ does not exceed the first critical value, $\mathcal{B} < \mathcal{B}_{c1}\sim m^2_B/g$, the scalar condensate does not depend on $\mathcal{B}$. At $\mathcal{B} = \mathcal{B}_{c1}$, the vacuum develops a raising gauge field condensate which inhibits the scalar condensate. The vacuum can experience two transitions in the form of smooth crossovers. 
% and define the vacuum instability at some critical values of magnetic fields.
The strings are energetically preferable field configuration when $\mathcal{B} > \mathcal{B}_{c1}\simeq (g/\alpha)\lambda\times 10^{20}\, T$  corresponding to the flux $\sim 2\pi/g$ inside the area $\sim 1/m_B^2$. The field $ E_z(r)$ is expelled from the vacuum and hence confined inside the region $r < m_B^{-1}$ that means the vortex-type, i.e., the configuration of the stringlike FT. The attractive forces can appear between the FTs and clustered into the CCOs. 
The further increasing of $\mathcal{B}$ meets the energy barrier to string formation, which vanishes within the energy balance when $\mathcal{B}_{b}\sim (\alpha/\sqrt\lambda ) \phi^2_0$ and we have the string formation at the distances $r\sim 1/m_B\sim 1/m_\varphi$. 
It means the balance of the propagation ranges of the gauge field and the scalar field.
For the magnetic fields which are away from the balance value $\mathcal{B}_b$, the interaction range of vector and scalar fields loses its balance, $g= \alpha/\sqrt\lambda$, and the FT interaction manifestly appears. 
At $\mathcal{B} = \mathcal{B}_b \simeq \sqrt\lambda \times 10^{20}\, T$, there is no interaction between the flux tubes.
%It means the balance of the propagation ranges of the gauge field and the scalar field.
%The $U(1)$ symmetry is restored locally. 
%The symmetry is restored totally 
When $\mathcal{B}$ achieves the second (transition) value $\mathcal{B} \sim \mathcal{B}_{c2} = m_\varphi^2/g$ at distances between the strings $r\sim 1/m_B >>1/m_\varphi$, one can use the mean field approximation (MFA)  when ${\vert\phi\vert}^2 = \phi_0^2$ with the cutoff $m_B = m_B \theta (r - m_\varphi^{-1})$.
%Here, $\alpha > \zeta_\phi\times 10^{-30} >> g > \sqrt\zeta_\phi \times 10^{-49}$, $\lambda\leq O(1)$.
%which may be much larger than the string core. 
The flux tubes repel each other,
% Here, $\}_B\alpha\sim \lambda\zeta_\phi (1- \Delta_B)^{-1}\times 10^{-34}$ with $\Delta_B\sim \zeta_B\times 10^{- 62} eV^2 < \zeta_B\times (10^{-24} - 10^{-20}$ and 
%the dark gauge coupling constant should be as $g_D < \zeta_\phi\times 10^{-34}$. 
%The scalar condensate should vanish and 
the FTs disappear entirely as the vacuum crosses into $\mathcal{B} > \mathcal{B}_{c2}\simeq (\alpha/g)\times 10^{20}\, T$, where the symmetry gets restored.
The critical values for magnetic fields $\mathcal{B}_{c1}$ and $\mathcal{B}_{c2}$ depend on the Ginzburg-Landau parameter $k_{GL} = \alpha/g = \sqrt\lambda \xi/l$ which defines the vacuum properties during the formation of the FT corresponding to the magnetic flux (\ref{e149}).

The $\bar B(r)$ and the $E_z (r)$ in the MFA  asymptotically behave as
 \begin{equation}
\label{e19}
\bar B(r)\simeq -\frac{1}{2\pi}\Phi - \sqrt{m_B r} e^{- m_B r},\,\,\,\   E_z(r)\simeq \frac{m_B^{3/2}}{\sqrt r} e^{- m_B r}\left ( 1 - \frac{1}{2 m_B r}\right )
\end{equation}
 %\begin{equation}
%\label{e19}
 %E_z(r)\simeq \frac{m_B^{3/2}}{\sqrt r} e^{- m_B r}\left ( 1 - \frac{1}{2 m_B r}\right )
%\end{equation}
at large $\xi\sim r > m_\varphi^{-1}$.
% (compared to that of the coherent length of the scalar field, $l\sim m_\varphi^{-1}$)
% with the replacement $\varphi (r) = \phi_0$ and the boundary condition $\bar B(r\rightarrow \infty) = - (1/2\pi)\Phi$. 
%The MFA assumes the FT can evolve in size and the mass, restricted by the cutoff  $m_B = m_B \theta (r - l)$. 
 The MFA breaks down at the core region of the FT. The cosmological compact object topologically corresponds to open FT excitations with the end points, the "terminals". 
%For $\zeta_\varphi\sim 10^{-3}$ (see, e.g., [7]) and $\lambda \leq O(1)$, the vector boson mass $m_B < m_\varphi \sim 10^{-7}$ eV that is much far away of the lower bound $\sim %2.2\cdot 10^{-21}$ eV (CL $> 95\%$) to date on the DM mass (independent of the spin nature of the fields) in the local Universe [8]. This means that 
Using the fact that the parameter $\alpha$ is set to be very tiny, 
%$\alpha\sim O(10^{-34})$ with $\zeta_\phi\sim O(1)$ 
in order to get the correct hierarchy rate (\ref{e1462}), the radial size of the FT is bounded by $ r\geq$ 20 cm with $\zeta_\phi\sim O(1)$.
% and 
%this $\xi\sim r$ defines the penetration depth of the $B_\mu$ field. 
%For the weak scalar-GR coupling, the radial 
%this size may increase drastically, $r > 6$ m, if $\zeta_\phi\sim 10^{-3}$, that has been used for a successful description of inflation in [10]. 
The flux  (\ref{e149}) is unconfined inside the region $r >> m_B^{-1}$ which means the FT configuration is deformed, becomes bloated, the scalar fields surrounding the tubes are distributed homogeneously inside the volume of the compact object. 
The latter can also be regarded as the massive astronomical FT (MAFT) with no "terminals", the co-called "the MAFT ring" excitation, where the symmetry is approximately restored.

%\section{Conclusion and discussion}
In conclusion, we studied the formation of the inhomogeneities in terms of the string like flux tubes imposed by the DM and the SM sectors minimally coupled to gravity. 
%DM scalar and vector fields interacting with the SM fields, which contribute to the formation of the string like objects. 
%The latter may be explained in terms of the fluxes of $B_\mu$ field under the impact of the complex scalar $\phi$ field as the dynamical quantity minimally coupled to GR. 
There are two stages of the EW vacuum in the background of magnetic field related to the formation of the FT at $r  < 1/m_B$ (the strings).
% and $r >>1/m_\varphi$ (MAFT, the symmetry is approximately restored). 
As the magnetic field reaches the first critical point $\mathcal{B}_{c1}$, the vacuum turns into the intermediate inhomogeneous phase, characterised by the presence of classical vector boson condensate.  
%The energy barrier between these stages vanishes when the balance between the coupling constants are achieved, $\alpha/\sqrt\lambda = g$.
At $\mathcal {B} = \mathcal {B}_{c2}$, the DM scalar condensate should vanish and the inhomogeneities disappear. 
The energy barrier between $ \mathcal {B}_{c1}$ and $\mathcal {B}_{c2}$ vanishes when 
%the balance between the coupling constants are achieved, 
$k_{GL} = \sqrt\lambda \xi/l = 1$.
%$\alpha/\sqrt\lambda = g$.
%The scalar field associated with the condensate may be difficult to detect as it has the vacuum quantum numbers. , the inhomogene
The important role of the scalar fields can be seeing when the scale invariance breaking disappears if the scalar fields are removing from the vacuum. 
This means the stability of the CCO and the scale symmetry breaking are strongly correlated to each other via the scalar sector.
The observation of the CCOs  may extend the spectrum of research reconsidering a few tens of years exploration to discover the scalar hidden sector as well as the  astrophysical objects,  composed of the fields from this sector as possible candidates to DM. 
One can admit, if the scalar DM gravitationally clustered into the compact object, this displays some universality with the SM sector. 
This can be seeing through the decays of the scalar fields because of instability of the compact object.
In some sense, the solutions with the profiles for $\bar B(r)$ and the field $E_z(r)$ in the wide range of $r$ is one of the best theoretical instruments for exploring the GR-potential effects 
such as the geometry of the FTs with "terminals" or the FT rings (with no "terminals"). The CCO binaries with repel forces between the FTs would be the ideal candidates to distinguish them from the binary black holes events to interpret the GW  signals detected to date [26].

\end{document}